\documentclass[twoside, twocolumn, 9pt]{article}
\usepackage{extsizes}
\usepackage[super,sort&compress,comma]{natbib} 
\usepackage[version=3]{mhchem}
\usepackage[left=1.5cm, right=1.5cm, top=1.785cm, bottom=2.0cm]{geometry}
\usepackage{balance}
\usepackage{amssymb}
\usepackage{sectsty}
\usepackage{graphicx} 
\usepackage{lastpage}
\usepackage[format=plain,justification=justified,singlelinecheck=false,font={stretch=1.125,small,sf},labelfont=bf,labelsep=space]{caption}
\usepackage{float}
\usepackage{fancyhdr}
\usepackage{fnpos}
\usepackage[english]{babel}
\addto{\captionsenglish}{%
  
}
\usepackage{array}
\usepackage{droidsans}
\usepackage{charter}
\usepackage[T1]{fontenc}
\usepackage[usenames,dvipsnames]{xcolor}
\usepackage{setspace}
\usepackage[compact]{titlesec}
\usepackage{hyperref}
\usepackage{bm}
\usepackage{epstopdf}%This line makes .eps figures into .pdf - please comment out if not required.

\definecolor{cream}{RGB}{222,217,201}

\begin{document}

\pagestyle{fancy}
\thispagestyle{plain}
\fancypagestyle{plain}{
%%%HEADER%%%
\renewcommand{\headrulewidth}{0pt}
}
%%%END OF HEADER%%%

%%%PAGE SETUP - Please do not change any commands within this section%%%
\makeFNbottom
\makeatletter
\renewcommand\LARGE{\@setfontsize\LARGE{15pt}{17}}
\renewcommand\Large{\@setfontsize\Large{12pt}{14}}
\renewcommand\large{\@setfontsize\large{10pt}{12}}
\renewcommand\footnotesize{\@setfontsize\footnotesize{7pt}{10}}
\renewcommand\scriptsize{\@setfontsize\scriptsize{7pt}{7}}
\makeatother

\renewcommand{\thefootnote}{\fnsymbol{footnote}}
\renewcommand\footnoterule{\vspace*{1pt}% 
\color{cream}\hrule width 3.5in height 0.4pt \color{black} \vspace*{5pt}} 
\setcounter{secnumdepth}{5}

\makeatletter 
\renewcommand\@biblabel[1]{#1}            
\renewcommand\@makefntext[1]% 
{\noindent\makebox[0pt][r]{\@thefnmark\,}#1}
\makeatother 
\renewcommand{\figurename}{\small{Fig.}~}
\sectionfont{\sffamily\Large}
\subsectionfont{\normalsize}
\subsubsectionfont{\bf}
\setstretch{1.125} %In particular, please do not alter this line.
\setlength{\skip\footins}{0.8cm}
\setlength{\footnotesep}{0.25cm}
\setlength{\jot}{10pt}
\titlespacing*{\section}{0pt}{4pt}{4pt}
\titlespacing*{\subsection}{0pt}{15pt}{1pt}
%%%END OF PAGE SETUP%%%

%%%FOOTER%%%
\fancyfoot{}
\fancyfoot[LO,RE]{\vspace{-7.1pt}\includegraphics[height=9pt]{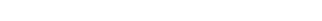}}
\fancyfoot[CO]{\vspace{-7.1pt}\hspace{13.2cm}\includegraphics{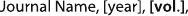}}
\fancyfoot[CE]{\vspace{-7.2pt}\hspace{-14.2cm}\includegraphics{head_foot/RF}}
\fancyfoot[RO]{\footnotesize{\sffamily{1--\pageref{LastPage} ~\textbar  \hspace{2pt}\thepage}}}
\fancyfoot[LE]{\footnotesize{\sffamily{\thepage~\textbar\hspace{3.45cm} 1--\pageref{LastPage}}}}
\fancyhead{}
\renewcommand{\headrulewidth}{0pt} 
\renewcommand{\footrulewidth}{0pt}
\setlength{\arrayrulewidth}{1pt}
\setlength{\columnsep}{6.5mm}
\setlength\bibsep{1pt}
%%%END OF FOOTER%%%

%%%FIGURE SETUP - please do not change any commands within this section%%%
\makeatletter 
\newlength{\figrulesep} 
\setlength{\figrulesep}{0.5\textfloatsep} 

\newcommand{\topfigrule}{\vspace*{-1pt}% 
\noindent{\color{cream}\rule[-\figrulesep]{\columnwidth}{1.5pt}} }

\newcommand{\botfigrule}{\vspace*{-2pt}% 
\noindent{\color{cream}\rule[\figrulesep]{\columnwidth}{1.5pt}} }

\newcommand{\dblfigrule}{\vspace*{-1pt}% 
\noindent{\color{cream}\rule[-\figrulesep]{\textwidth}{1.5pt}} }

\makeatother
%%%END OF FIGURE SETUP%%%

%%%TITLE AND AUTHORS%%%
\twocolumn[
  \begin{@twocolumnfalse}
{\includegraphics[height=30pt]{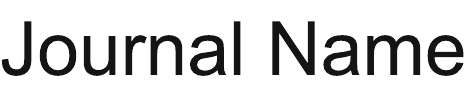}\hfill\raisebox{0pt}[0pt][0pt]{\includegraphics[height=55pt]{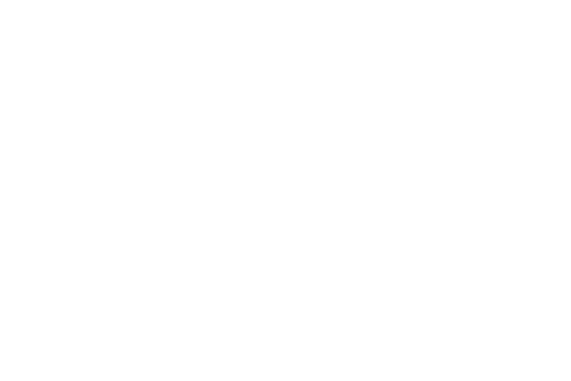}}\\[1ex]
\includegraphics[width=18.5cm]{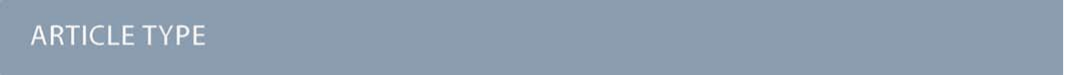}}\par
\vspace{1em}
\sffamily
\begin{tabular}{m{4.5cm} p{13.5cm} }
\includegraphics{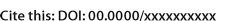} & \noindent\LARGE{\textbf{Quantum and Critical Casimir Effects: Bridging Fluctuation Physics and Nanotechnology}} \\%Article title goes here
 & \vspace{0.3cm} \\
 & \noindent\large{Roberto Passante,\textit{$^{a, b}$} Lucia Rizzuto,\textit{$^{a, b}$} Peter Schall,\textit{$^{c}$} and Emanuele Marino$^{\ast}$\textit{$^{a}$}} \\%Author names go here instead of "Full name", etc.
\includegraphics{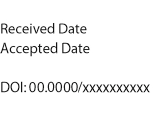} & \\
\end{tabular}
 \end{@twocolumnfalse} \vspace{0.6cm}
  ]
%%%END OF TITLE AND AUTHORS%%%

%%%FONT SETUP - please do not change any commands within this section
\renewcommand*\rmdefault{bch}\normalfont\upshape
\rmfamily
\section*{}
\vspace{-1cm}

%%%FOOTNOTES%%%

%\footnotetext{\textit{$^{a}$~Address, Address, Town, Country. Fax: XX XXXX XXXX; Tel: XX XXXX XXXX; E-mail: xxxx@aaa.bbb.ccc}}
%\footnotetext{\textit{$^{b}$~Address, Address, Town, Country. }}
\footnotetext{\textit{$^{\ast}$ E-mail: emanuele.marino@unipa.it}}
\footnotetext{\textit{$^{a}$ Department of Physics and Chemistry - Emilio Segr\`{e}, University of Palermo, Via Archirafi 36, 90123 Palermo (PA), Italy.}}
\footnotetext{\textit{$^{b}$ Istituto Nazionale di Fisica Nucleare, Laboratori Nazionali del Sud, I-95123 Catania, Italy}}
\footnotetext{\textit{$^{c}$ Van der Waals-Zeeman Institute of Physics, University of Amsterdam, Science Park 904, 1098XH Amsterdam, The Netherlands}}
%%%END OF FOOTNOTES%%%

%%%ABSTRACT%%%%

\sffamily{ 
\section*{Abstract}\textit{Fluctuation-induced forces, primarily represented by quantum and critical Casimir effects, play a pivotal role at the nanoscale. This review explores the theoretical and experimental landscapes of these forces, offering a comprehensive analysis of their similarities and distinctions. We emphasize the effects of material properties, geometry, and temperature in shaping these forces and their roles in various nanoscale systems, both colloidal and solid-state. We devote special attention to the Casimir torque, the influence of magnetism on the Casimir force, and the use of Casimir effects for the generation of optical resonators. Through this comparative study, we elucidate the underlying physics of these phenomena, fostering insights that advance applications in nanomechanics, optomechanics, and quantum technologies.\\\\}
}

\rmfamily %Please do not remove this line.

%%%MAIN TEXT%%%%

% TO DO 
% 1. Derive in some way the critical Casimir effect following Peter's review 
% 2. Write conclusions 
% 3. Remove simplified formula 

\section{Introduction}

%%%%%%%%%%%%% Fluctuations
\subsection{Fluctuations and fluctuation-induced interations}\label{sec:fluctuation}
Fluctuations can be defined as deviations of the value of an observable from its average. As a consequence of the generality of the fluctuation-dissipation theorem,\cite{Callen_Welton51} fluctuations are ubiquitous to many different areas of physics, from statistical mechanics to condensed matter physics, quantum field theory, and cosmology.\cite{Intravaia-Henkel11} Perhaps unexpectedly, the spatial confinement of field fluctuations between two macroscopic bodies yields a long-range force known as \textit{fluctuation-induced force}. The \textit{quantum Casimir effect} represents a prominent example of a force induced by fluctuations.\cite{Casimir48, Bimonte_Thorsten22}

%%%%%%%%%%%%% Casimir effect
\subsection{Casimir Effect: Quantum versus Critical} 
The quantum Casimir effect arises from the confinement of the quantum vacuum fluctuations of the electromagnetic field between two uncharged macroscopic objects, even at zero temperature. The presence of boundary conditions restricts the allowed quantum modes of the electric ($\boldsymbol{E}$) and magnetic ($\boldsymbol{B}$) fields, resulting in a net force acting on the confining bodies. In the case of two electrically neutral, perfectly conductive and parallel plates separated by a distance $L$, the quantum Casimir force is attractive and its magnitude decays as $L^{-4}$. Short-range corrections to this behavior are present in the case of plates with finite conductivity or in the presence of a dielectric. \cite{Bordag-Klimchitskaya09,KlimchitskayaMohideen2009,BrevikShapiro22}.

%\subsection{The Critical Casimir Effect}
Moving from quantum to classical fluctuations yields a conceptually similar effect known as the \textit{critical Casimir effect}. When a fluid is brought close to its critical point, solvent density fluctuations arise. These fluctuations can be described in terms of the deviation of the local mass density from the average, $\partial c(\boldsymbol{x}) \equiv c(\boldsymbol{x}) - c$, where $\boldsymbol{x}$ indicates the position vector. Bringing the system closer to the critical point increases the size of these fluctuations, formalized as their spatial correlation length $\xi$, defined as $\langle\partial c(\boldsymbol{x})\partial c(\boldsymbol{x\prime})\rangle - \langle\partial c(\boldsymbol{x})\rangle\langle\partial c(\boldsymbol{x\prime})\rangle\propto exp{(-|\boldsymbol{x}-\boldsymbol{x\prime}|/\xi)}$. The confinement of these fluctuations between two macroscopic objects results in a net force acting on the confining bodies known as the critical Casimir force.\cite{DeGennes78, Gambassi09, Dantchev23, Dantchev24} Like the quantum Casimir effect, the critical Casimir effect obeys a power-law behavior; in the case of two parallel plates, the magnitude of the critical Casimir force decays as $L^{-3}$. 

%%%%%%%%%%%%% Scaling arguments
\subsection{Scaling Arguments}
The power-law behaviors observed for quantum ($L^{-4}$) and classical ($L^{-3}$) Casimir forces can be qualitatively inferred through dimensional analysis. For the geometrical configuration of two parallel plates, the force per unit area is given by $\mbox{Force/Area} \sim  \mbox{Energy/Length}^{3}$. Since for a quantum system $\mbox{Energy}\sim hc/\lambda$, we obtain for the quantum Casimir effect $\mbox{Force/Area}\sim \mbox{Length}^{-4}$. Instead, for a classical system $\mbox{Energy}\sim k_B T$, resulting for the critical Casimir effect in the scaling law $\mbox{Force/Area}\sim \mbox{Length}^{-3}$. 

\subsection{This Review}
In this paper, we review and discuss key features of both quantum and critical Casimir effects, their analogies and differences, focusing on their relevance to the development of future technologies. 
We first review the quantum Casimir effect from a theoretical point of view, evaluating explicitly the force between two perfect plane mirrors in vacuum (section \ref{sec:qCas}).
Then, we derive the critical Casimir effect and discuss analogies and differences with the quantum effect (section \ref{sec:cCas}).% To be addressed

Furthermore, we explore how the Casimir effects - both quantum and critical - are being used today as driving force in the development of new optoelectronic devices operating at the nanoscale. We begin by exploring how Casimir torques use rotation to achieve alignment between optically-responsive materials (section \ref{sec:torques}). We continue exploring the role of quantum and critical Casimir forces in  understanding adhesion between solids (section \ref{sec:SSCas}) and colloids (\ref{sec:CollCas}), enabling the generation of self-assembled optical resonators to trap light at the nanoscale. 

Finally, we discuss recent efforts in achieving tunable Casimir forces either by controlling the dielectric environment (section \ref{sec:TunCasDie}) or by introducing an external tunable bias, such as a magnetic field (section \ref{sec:TunCasMag}).
We conclude by providing a perspective on this fascinating field. 

%%%%%%%%%%%%% Quantum Casimir effect
\subsection{The quantum Casimir effect}\label{sec:qCas}
\subsubsection{Background}
Since the theoretical prediction by Casimir in 1948,\cite{Casimir48} the quantum Casimir effect has been extensively investigated. However, to this day measuring quantum Casimir forces continues to represent a significant technical challenge because of the small magnitude of the interaction. In the ideal case of two plane perfectly reflecting mirrors of area $\sim 1 \, cm^2$ separated by a distance of $\sim 1 \, \mu m$, the quantum Casimir force is attractive and of the order of $\sim 10^{-7}\, N$. Measuring such weak forces has become possible only recently, first through an electromechanical torsion pendulum\cite{PhysRevLett.78.5} and then through atomic force microscopy.\cite{Munday2009}. These measurements have allowed the community to chase even finer aspects of the Casimir effect. Indeed, when the bodies confining quantum vacuum fluctuations feature a high degree of anisotropy, such as two parallel birefringent plates or gratings, theory predicts that a momentum of the quantum Casimir force - \textit{a Casimir torque} - should develop to minimize the energy of the system through rotation.\cite{PhysRevLett.124.013903} The first measurements of the Casimir torque relied on an optical setup to measure the alignment of a liquid crystal to a birefringent crystal in real time, resulting in experimental values of the Casimir torque per unit area of the order of $\sim 10^{-8}\, N/m$,\cite{Somers2018} prompting the development of more sensitive detectors in the last 5 years.\cite{Ahn2020}

%%%%%%%%%%%%% Technological Relevance of the Quantum Casimir effect
Despite their small magnitude, quantum Casimir forces hold crucial relevance for nanotechnology in the fabrication of current and future electromechanical devices operating at the nanoscale. At that length scale, system design must prevent device failure by the permanent adhesion ({\em stiction}) between device components due to the quantum Casimir force. Furthermore, novel designs rely actively on the quantum Casimir force to construct extremely sensitive force sensors \cite{Woods-Dalvit16}, {\em Casimir machines} for moving microscopic objects.\cite{Stange} Even the experimental realization of a Casimir diode\cite{Xu2022NonReciprocal} and Casimir transistor\cite{Xu2022} have recently been investigated. These works suggest that the Casimir effect may be exploited to fabricate a new class of devices with novel properties. Therefore, understanding, controlling, and exploiting the Casimir effect in different material systems and configurations is of paramount interest to future technological developments. 

%%%%%%%%%%%%% Boundary effects
Boundary conditions play a crucial role in defining the Casimir interaction, since the presence of boundaries affects the vacuum fluctuations of the electromagnetic field. Therefore, the quantum Casimir effect is very sensitive to the magneto-dielectric properties of the confining bodies, highlighting the central role of material choice and control. Lifshitz was the first to attempt a generalization of the Casimir effect; for this reason, the Casimir effect generalized to different materials and temperatures is known as the \textit{Casimir-Lifshitz effect}.\cite{Lifshitz} More specifically, quantum Casimir interactions can be inhibited or enhanced by material choice. For instance, recent literature reports have shown that the attractive Casimir force between two surfaces can turn repulsive when choosing the dielectric properties of the system appropriately\cite{Munday2009, Zhao2019} or considering bodies with non trivial geometry \cite{Zietal}. Besides dielectric control of the experimental system, the exploitation of magnetic fields to tune the Casimir force from attractive to repulsive has shown promise.\cite{Casimir-magnetic-experiment} However, achieving direct control over the interaction continues to represent an experimental and theoretical challenge in spite of the advances in manipulating the Casimir force.\cite{Casimir-Quantum-Plus-Critical}

 %%%%%%%%%%%%% Non additive interactions
One of the reasons behind the elusive control of Casimir forces may be their non-additivity.\cite{Present1971}. Interactions are said to be additive if the interaction between any two bodies of an ensemble of $N$ is not influenced by the presence of the remaining $N-2$ bodies. In this case, the total interaction can be represented as a sum of $N(N-1)/2$ pair interactions. Yet, the Casimir effect is sensitive to the morphology and relative (time-dependent) arrangement of all field-confining bodies, leading to non-additive interactions. This is true for all manifestations of the effect, both quantum - Casimir-Lifshitz\cite{Jeyar23} and Casimir-Polder\cite{Passante05, Fuchs_Bennett18, Milton_Abalo15} - and classical - critical Casimir.\cite{Paladugu}

\subsubsection{Derivation of the quantum Casimir force for plate-plate interactions}
\begin{figure}[!h]
\centering
  \includegraphics[width=0.7\columnwidth]{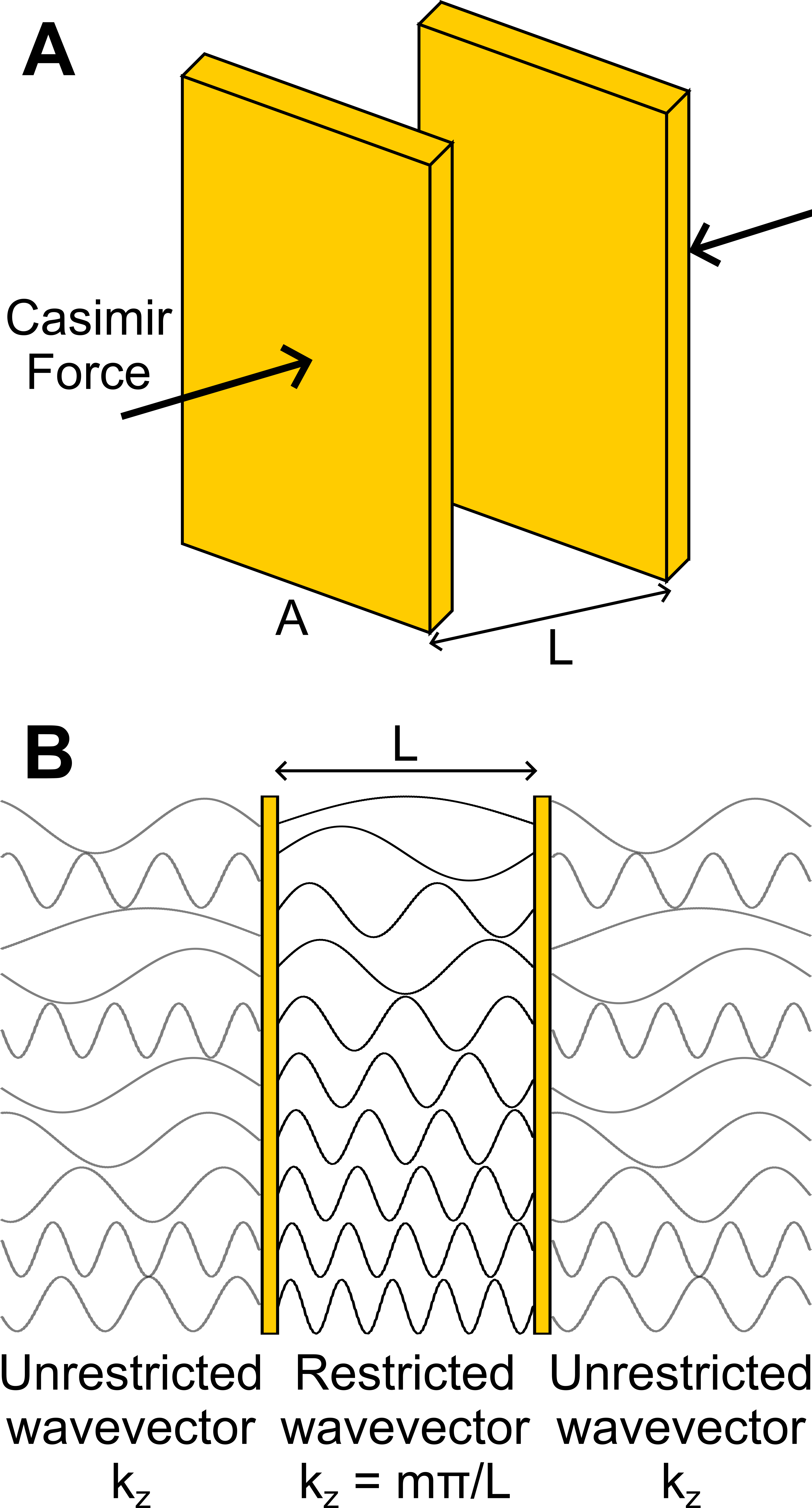}
  \caption{\textbf{(A)} Two perfectly conducting plates plates of area $A$ separated by a distance $L$ in the quantum vacuum. \textbf{(B)} The boundary conditions impose a limit on the allowed frequencies between the plates.}
  \label{fig1}
\end{figure}
The theory of quantum electrodynamics shows that, even at zero temperature, the electric and magnetic fields possess fluctuations such that the energy of the ground state of the electromagnetic field - that is, in the absence of photons - can be written as: 
\begin{equation}
E_0=\sum_{\bf k}\hbar\omega_k,
\label{eq:2.1}
\end{equation}
where ${\bf k}$ spans over all wavevectors of the field modes allowed by  boundary conditions.
This expression diverges when considering a summation over all frequencies. However, the allowed frequencies $\omega_k$ of the system depend on the boundary conditions imposed on the fields; thus the zero-point energy changes if the boundary conditions change. This eventually yields a potential energy between them.

Let us consider two parallel and perfectly conductive uncharged plates of area $A$ in vacuum separated by a distance $L$, see Fig.\ref{fig1}A. Since the parallel component of the electric field must go to zero at the surface of the plates, the very presence of the plates imposes specific boundary conditions on the electric and magnetic fields (see Fig.\ref{fig1}B), such that the normal ($z$) component of the wavevector ${\bf k}$ can only take the discrete values
\[
k_z=\frac{m\pi}{L},
\]
where $m=0,1,2,..$, while $k_x$ and $k_y$ can assume continuous real values.  
Therefore, the zero-point energy of the field is given by:
\begin{eqnarray}
\label{eq:2.2}
E_0(L)=\frac{L^2}{\pi^2}\hbar c\sum_{m}\,^{'}\int dk_x\int dk_y \sqrt{k_x^2+k_y^2+(\frac{m\pi}{L})^2},
\end{eqnarray}
where the prime in the sum means that we must multiply by a factor $1/2$ the term with $m=0$, to take into account that for $k_z =0$ we have only one independent polarization. The quantity described by Eq. \ref{eq:2.2} diverges in any finite volume. If we now suppose that the plates are placed at infinite distance from each other by taking the limit $L\rightarrow\infty$, the zero-point energy assumes the form:
\begin{eqnarray}
\label{eq:2.3}
E_0(\infty)=\frac{L^3}{\pi^3}\hbar c\int dk_x\int dk_y \int dk_z\sqrt{k_x^2+k_y^2+k_z^2},
\end{eqnarray}
that diverges too. However, we can consider that the physically meaningful quantity is the potential energy related to the work done to bring the plates from an infinite to a finite separation $L$, that is:
\[
U(L)=E_0(L)-E_0(\infty).
\]

After a regularization procedure, this eliminates the divergences and yields a finite quantity depending on the distance $L$ between the two plates, the Casimir energy:
\begin{eqnarray}
\label{eq:2.3}
U(L)=E_0(L)-E_0(\infty)=-\frac{\pi^2\hbar c A}{720 L^3}
\end{eqnarray}
The attractive force per unit area between the plates (or, equivalently, the pressure) can be then obtained in a quasi-static limit as the spatial derivative of Eq. \ref{eq:2.3} yielding: 
\begin{eqnarray}
\label{eq:2.4}
\frac{F(L)}{A}= - \frac{\pi^2\hbar c}{240 L^4}.
\end{eqnarray}
We notice that this expression depends only on the distance $L$ and the two fundamental constants $\hbar$ and $c$, as we have anticipated in section \ref{sec:fluctuation} based on dimensional arguments. The expression in Eq. (\ref{eq:2.4}) does not depend on the electric charge $e$, as a consequence of the assumption of perfectly conductive plates that reflect all the incident light. 

A more general calculation of the Casimir interaction between planar dielectrics, using the dielectric function of the bodies, was provided by Lifshitz in $1956$.\cite{Landau-Pitaevskii84, Milonni04} In the case of two semi-infinite dielectric slabs with dielectric permittivities $\epsilon_1(\omega )$ and $\epsilon_2 (\omega )$ separated by a distance $d$ in vacuum $(\epsilon_3(\omega ) = \epsilon_3=1)$, and in thermal equilibrium with the environment at temperature $T$, the Casimir force obtained by the Lifshitz formula is
%AGGIUNGERE FIGURA DELLA CONFIGURAZIONE
\begin{eqnarray}
\label{eq:1}
&\ &F(d) =\frac{k_B T}{\pi c^3}\sum_{n=0}^{\infty}\prime\,\,\xi_{n}^{3}\int_{1}^{\infty}dp p^2\Biggl\{\biggl [\frac{s_{1n} + \epsilon_{1n}p}{s_{1n} - \epsilon_{1n}p}\frac{s_{2n} + \epsilon_{2n}p}{s_{2n} - \epsilon_{2n}p}\nonumber\\
&\ &\times e^{2p\xi_n d/c}-1\biggr]^{-1}+\biggl [\frac{s_{1n} + p}{s_{1n} - p}\frac{s_{2n} + p}{s_{2n} - p}e^{2p\xi_n d/c}-1\biggr]^{-1}\Biggr\}
\end{eqnarray}
where $k_B$ is Boltzmann's constant, $\epsilon_1$ and $\epsilon_2$ the dielectric permittivities evaluated at the imaginary frequency $\omega=i\xi$ and the prime in the summation below means that we must multiply with $1/2$ when $n=0$. The expressions 
\begin{eqnarray}
\label{eq:2}
s_{jn}=\sqrt{p^2-1+\epsilon_{jn}},\,\,\, \mbox{with}\,\,\,\epsilon_{jn}=\epsilon_j(i\xi_n)\,\, (j=1,2),
\end{eqnarray}
and 
\begin{equation}
\xi_n=\frac{2\pi k_B T}{\hbar}n \qquad n=0,1,2,...
%hbar does not show
\end{equation}
represent the Matsubara frequencies.
The expression in Eq. (\ref{eq:1}) is very general and many applications of this theory to planar multilayered dielectric media have been proposed in the literature \cite{Bordag-Klimchitskaya09,WoodsDalvit16}, including nonlocal spatial response \cite{KlimchitskayaMostepanenko21}.
However, in general an analytical calculation of the Casimir interaction can be very complicated, and some approximations have been proposed in the literature to generalize the theory to complex non-planar geometries \cite{Intravaia-Koev13}, allowing to evaluate the Casimir force in more general situations, with objects of complex shape and real materials.
\cite{Johnson11, Parsegian05}. 
As anticipated in section \ref{sec:qCas}, in these cases the expression of the Casimir force can be very different with respect to the case of two ideal plates placed in vacuum. 
 
%%%%%%%%%%%%% Critical Casimir effect
\subsection{The critical Casimir effect}\label{sec:cCas}
\subsubsection{Background}

%Intro
In 1978, thirty years after Casimir's prediction, Fisher and de Gennes postulated that a classical effect similar to the quantum Casimir effect should be observed in a fluctuating physical medium, such as a mixture of two fluids close to the critical point.\cite{DeGennes78} Near criticality, extended regions with fluctuating density (or composition) emerge in the bulk of the solvent; Fisher and de Gennes understood that fluctuation-induced forces - critical Casimir forces - should arise when these solvent density fluctuations are spatially confined between two immersed bodies. Similarly to the quantum Casimir force, the critical Casimir force is relatively weak: for example, at room temperature ($T \sim 300 \,K$) and for plates of area $1 \, cm^2$ separated by a distance of $1 \,\mu m$, the critical Casimir force is of the order of  $10^{-7} \,N$, comparable with the quantum Casimir force at zero temperature.\cite{Gambassi09}.

%Description
The critical Casimir force between a sphere and a plate immersed in a binary fluid mixture close to its critical point was measured for the first time only in 2008 by using total internal reflection microscopy.\cite{Hertlein2008} In this technique, the amount of light captured by a microscope objective depends sensitively on the distance of the colloidal particle from the plate, allowing the extraction of the distribution of particle-wall distances $P(d)$ and hence - by inverting Boltzmann's equation - the particle-wall potential $U(d)$.\cite{PRIEVE199993} Since then, the use of confocal microscopy has enabled the observation of the assembly of spherical colloids,\cite{PhysRevLett.103.156101} and the extraction of their pair potential.\cite{Nguyen2013} As predicted by theory and shown by experiments, the magnitude and length scale of the critical Casimir force depend sensitively on the size (correlation length, $\xi$) of the solvent density fluctuations, which in turn depend sensitively on how far the system is from criticality in terms of composition $c$ and temperature $T$ of the mixture. Section \ref{sec:critCasTheory} provides a brief theoretical background on the critical Casimir effect.

\begin{figure}[!h]
\centering
  \includegraphics[width=1\columnwidth]{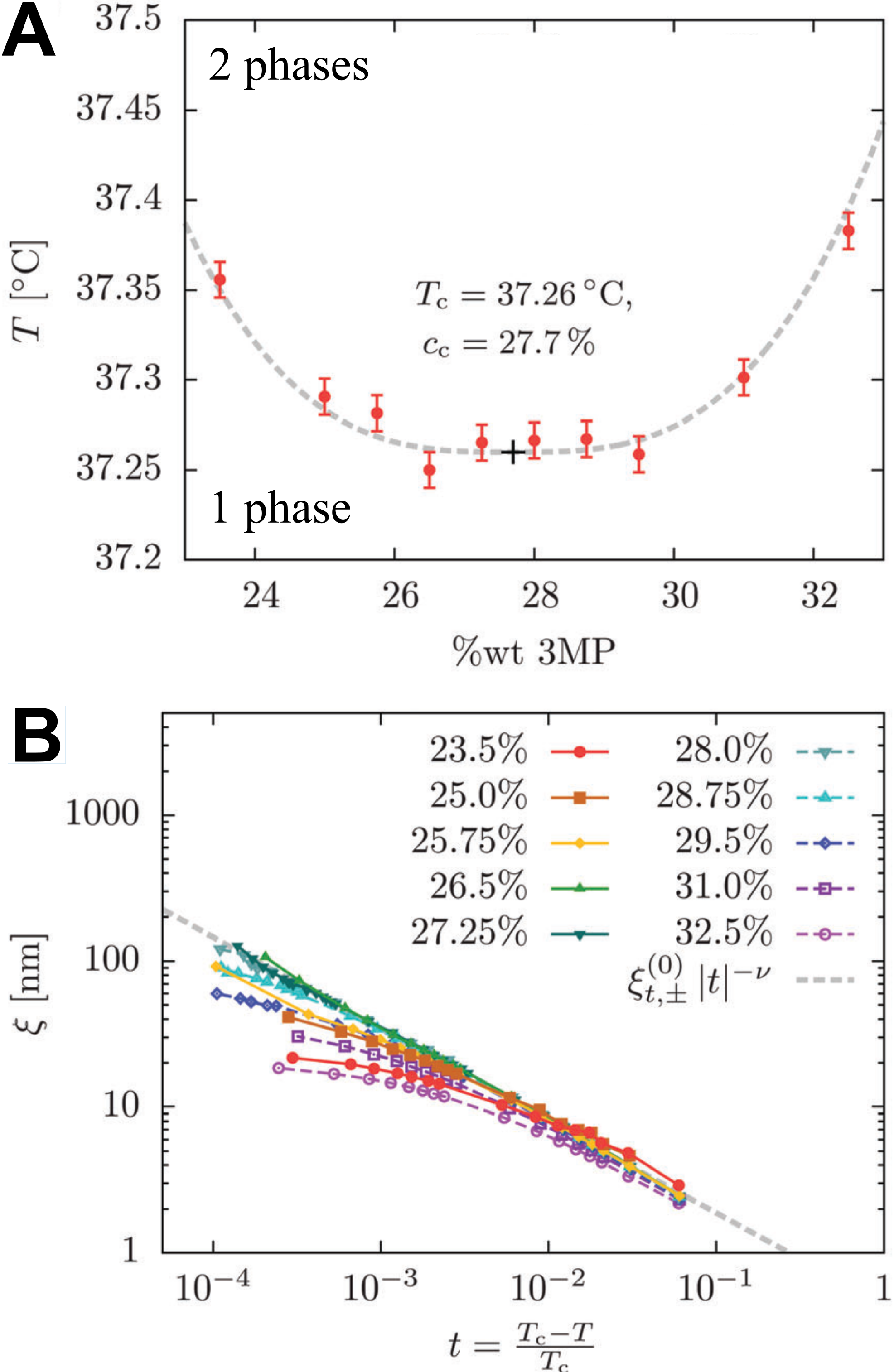}
  \caption{\textbf{Phase diagram of a binary mixture of heavy water and 3-methylpyridine (3MP) and correlation length determined by dynamic light scattering.} 
  \textbf{(A)} Phase separation temperature as a function of the weight fraction of 3MP in the mixture. Data points were determined by dynamic light scattering while dashed line describes the behavior expected from theory. The critical point is located at the coordinates at $T_c = 37.26\,^\circ C$ and $c_c = 27.7\,w/w\,\%$.  
  \textbf{(B)} Correlation length $\xi$ as a function of $(T_c - T)/T_c$ for various concentrations of 3MP ($w/w\,\%$). Near the critical composition, the correlation length follows the power-law expected from theory $\xi(T) = \xi_0|(T_c-T)/T_c|^{-\nu}$ with $\xi(0) = 0.44\,nm$.
  Panels A and B are adapted with permission from ref. \cite{C7SM00599G}. Copyright 2017 The Royal Society of Chemistry.
  }
  \label{fgr:phase}
\end{figure}

%Phase diagrams
Before proceeding further, the reader may find useful a brief introduction to the phase diagram of a binary solvent.\cite{LAW2001159} It is well known that two fluid components may be immiscible, such as water an oil, or miscible, such as water and alcohol. In this latter case, increasing (decreasing) the temperature of a well-mixed binary mixture may cause separation into two phases, one richer in the first component and the other richer in the second. The temperature at which this separation takes place is known as the \textit{phase separation temperature} $T_s$. Mapping the dependence of $T_s$ on the composition of the binary mixture $c$ - typically the weight or molar fraction of one component into the mixture - yields a convex (concave) curve. The minimum (maximum) of this curve is known as the lower (upper) critical solution temperature and corresponds to the critical point of the mixture identified by coordinates $T_c$, the critical temperature, and $c_c$, the critical composition. While most binary systems feature an upper consolute temperature (such as methanol and cyclohexane),\cite{MARHOLD1998127} binary systems characterized by a lower critical solution temperature (such as water and 2,6-dimethylpyridine (lutidine) or 3-methylpyridine (picoline, 3MP)),\cite{JR9520004606} are of more practical use in the laboratory and have been used for studies of the critical Casimir effect. For this reason, we will be referring only to binary mixtures with a lower critical solution temperature. Fig. \ref{fgr:phase}A shows the bottom portion of the experimental phase diagram of such a system, consisting of heavy water and 3MP.\cite{C7SM00599G} 

%Temperature

At the critical composition $c_c$ and for temperatures below the critical value $T_c$, theory and experiments agree on a power-law dependence $\xi(T) = \xi_0|(T_c-T)/T_c|^{-\nu}$, where $\xi_0$ is a solvent-specific value ($0.44\,nm$ for heavy water and 3MP),\cite{C7SM00599G} and $\nu = 0.63$ is the Ising critical exponent, see Fig. \ref{fgr:phase}B. As the temperature of the system approaches $T_c$, the critical Casimir potential becomes long-ranged ($\xi\gg d$), diverging at the critical point. Instead, decreasing the temperature far below the critical value results in a sharp decrease in the length scale of the interaction ($\xi\ll d$), effectively turning off the critical Casimir effect. Typical measurements have been performed several $10$'s to $100$'s of $mK$ below $T_c$, where significant critical Casimir forces have been observed. The thermal reversibility of the critical Casimir effect is one of its most interesting features,\cite{PhysRevLett.100.188303} and represents an important distinction from the quantum Casimir effect where the zero-point fluctuations of the electromagnetic field are always present.\cite{Casimir-Quantum-Plus-Critical} The thermal control of this interaction has enabled a number of experimental observations over the last two decades, such as temperature-controlled colloidal crystallization\cite{Nguyen2013, Kennedy_2022}, gelation,\cite{Rouwhorst2020, PhysRevLett.132.078203} and fractal formation,\cite{Shelke2013} the realization of colloidal analogues of molecular systems,\cite{Swinkels2021, Stuij2021} the observation of critical Casimir torques,\cite{Casimir-torque-disks, PhysRevE.88.012137,labbe2014alignment} and the observation of the non-additivity of critical Casimir forces.\cite{Paladugu, Magazzu2019, PhysRevE.91.042304, 10.1063/1.4791554}

%Composition
Moving away from the critical composition ($c\neq c_c$) influences the critical Casimir effect significantly. Experiment and theory agree that the largest critical Casimir effect should be expected on the side of the critical composition that is poor in the component preferred by the particles.\cite{10.1063/1.4722883} This is because the composition of the solvent density fluctuations is rich in the minority component in the bulk of the solvent. This consideration introduces the importance of boundary conditions in defining the critical Casimir effect. Given two confining surfaces, we can define symmetric ($++$ or $--$) or asymmetric ($+-$ or $-+$) boundary conditions depending on whether both surfaces preferentially adsorb to the same component of the mixture or have opposite preferences. As may be expected, symmetric boundary conditions lead to attractive critical Casimir interactions, while asymmetric boundary conditions lead to repulsive critical Casimir interactions\cite{RevModPhys.90.045001, PhysRevLett.101.208301}. This is another important difference with the quantum Casimir effect, where the interactions are primarily attractive, although recent works have shown that the Casimir force between a perfect electrically conducting plate and a perfect magnetically conducting plate is repulsive.\cite{Boyer74,BreviK18} More recently, the subtle interplay between quantum and critical Casimir effect has been exploited to balance the attractive Casimir force between microscopic gold flakes with repulsive critical Casimir forces.\cite{Casimir-Quantum-Plus-Critical}. 

%The protein limit
Most studies in the literature use colloids with a radius larger than the correlation length of the solvent density fluctuations, $R\gtrsim\xi$. The reasons behind this choice are mostly experimental: total internal reflection microscopy requires particles with a sufficiently large scattering cross section, while confocal microscopy requires particles significantly larger than the optical resolution. For these reasons, the regime $R<\xi$, known as the \textit{protein limit}, has remained largely unexplored. However, the recent development of solvothermal techniques for the synthesis of colloidal nanoparticles with controlled morphologies, compositions, and physical properties has motivated the community to the introduction of smaller and smaller colloidal systems. In 2016 we used dynamic light scattering to show that the critical Casimir effect can induce the assembly of colloidal semiconductor nanoparticles, also known as quantum dots.\cite{MARINO2016154} One of the most striking differences from operating outside the protein limit is the widening of the assembly region to temperatures as far as $20\,K$ from $T_c$. This experimental observation suggests that $\xi\sim R$ may be considered as a rule of thumb for the onset of colloidal assembly.\cite{Nguyen_2016} The introduction of colloids with optoelectronic functionality introduces the possibility of applications based on the critical Casimir effect. We demonstrated that the critical Casimir assembly can induce the formation of films with fractal morphology for field-effect transistors,\cite{Marino2019} with a larger degree of ordering when the assembly was performed in two dimensions.\cite{Marino2021, Vasilyev2021}

%T>Tc
Out of completeness, we now discuss the behavior of a binary fluid mixture above the critical temperature, $T>T_c$ for binary systems with a lower critical solution temperature as shown in Fig. \ref{fgr:phase}A. When working at the critical composition $c=c_c$, increasing the temperature of the fluid mixture above the critical temperature $T_c$ causes the decomposition of the mixture into two phases, each richer in one of the two fluid components. This decomposition shows a transient during which the morphology of the decomposition is spinoidal,\cite{Herzig2007} after which the fluid mixture fully separates into two components. At off-critical compositions $c\neq c_c$, mixture decomposition takes place by the nucleation of droplets rich in the minority component present in the bulk of the solvent. The absence of fluctuations in the two-phase region of the phase diagram restricts the critical Casimir effect to the one-phase region. Nevertheless, the two-phase region and phase separation processes have recently garnered the interest of part of the scientific community through the synthesis of bijels\cite{10.1063/5.0048797} and suspoemulsions.\cite{ZHENZHOU2024471}

\subsubsection{Derivation of the critical Casimir force for plate-plate interactions}\label{sec:critCasTheory}

The characteristics of the quantum Casimir effect discussed in section \ref{sec:qCas} are not exclusive to the quantum electromagnetic field. In fact, the key element underlying the quantum Casimir effect is the presence of a fluctuating field (the quantum electromagnetic field in that case) that can be modified by the presence of physical boundaries giving rise to a {\em configurational} energy that depend on the distance between the boundaries. Thus we can expect analogous effects in the presence of  fluctuations of different nature. This is the case of the critical Casimir effect.

To understand the physical origin of the critical Casimir effect, we consider two material bodies $A$ and $B$ immersed in a fluid  at room temperature. In such a system, fluctuations of the density of the fluid usually occur at the typical molecular scale $\xi_0$. There may be an effective force $F_{A,B}$  between the two bodies mediated by the fluid; if the thermodynamic state of the fluid is far away from a critical point, this force changes slowly with the temperature $T$.  However, near a critical point, characterized for example by specific values of the temperature, the fluctuations of the density of the fluid (or of some other relevant quantity in the system, usually known as order parameter of the phase transition) can become relevant even at scales $\xi$ much larger than the molecular length $\xi_0$. This is because, near a second-order phase transition, the collective behavior of the molecules in the fluid introduces a new correlation length in the system. The physical behavior of the system is then related to the fluctuations of the order parameter $\phi$, with a spatial correlation length $\xi$ that increases upon approaching the critical point. 

Analogously to the case of the quantum Casimir effect, when two external bodies are immersed in the fluid, they impose boundary conditions on the order parameter and modify the spectrum of its fluctuations and the free energy of the system, that now depend on the position of the two bodies.
This gives rise to an additional contribution to the force $F_{A,B}$ between the two bodies, which is the critical Casimir force $F^{Cas}$.  For two parallel plates of area $A$ , separated by a distance $L$, this force is given by \cite{Gambassi09, Dantchev23, Dantchev24}
\begin{eqnarray}
\label{eq:3.1}
\frac{F^{Cas}}{A}=\frac{k_B T}{L^3}\vartheta_{||}(L/\xi),
\end{eqnarray}
where $\vartheta_{||}(L/\xi)$ is a universal function based only on general properties of the critical behaviour of the system and to the specific boundary conditions imposed on the order parameter. The microscopic detail of the interaction between the molecules of the fluid and the dipoles in the plates are irrelevant to determine the force (assuming the correlation length $\xi$ and the distance $L$ much larger than the microsopic length scales of the system).
All the previous considerations can be applied also to the case of a mixture of two fluids approaching a critical point in the demixed phase, and confined by two boundaries that affect the allowed fluctuations and consequently the free energy of the system.\cite{Gambassi09}

The analogy between the distance scaling of the quantum Casimir force between perfectly reflecting parallel plates discussed in section \ref{sec:qCas} and that of the critical Casimir force in Eq.(\ref{eq:3.1}) shows in the latter the additional factor $\vartheta_{||}(L/\xi)$ of the dimensionless parameter $L/\xi$. This difference is related to the fact that in the critical case there is an intrinsic length scale given by the correlation length, while in the quantum electrodynamical Casimir force for ideal plates there is no relevant length scale because the photon is a massless particle and the boundaries are assumed ideal. The situation changes for the quantum Casimir force for a massive vector field characterized by a quanta mass $m$, where a new length scale emerges given by the Compton wavelength of the field quanta $\lambda_C=\hbar /mc$.\cite{Barton-Domey84,BARTON1985231,PLUNIEN198687,farina2006} In this case, the explicit calculation based on the Proca equation yields an analogous correction factor of the dimensionless parameter $L/\lambda_C$. When the distance between the plates is small, $L \ll \lambda_C$, the expression for this modified Casimir force reads:\cite{Barton-Domey84}. 

\begin{equation}\label{Cmp1}
\frac{F(L)}{A} \simeq - \frac{\pi^2\hbar c}{240 L^4} \left[ 1-5 \left( \frac {L}{\lambda_C}\right)^2 \right] ,
\end{equation}

\noindent showing a reduction of the Casimir force compared to the massless photon case, Eq.\ref{eq:2.4}. The quantity in the square brackets in Eq.\ref{Cmp1} can be considered as an approximated  scaling factor of the dimensionless ratio $L/\lambda_C$. Instead, when the distance between the plates is large, $L \gg \lambda_C$, an exponential factor $e^{-2L/\lambda_C}$ is present in the correction factor \cite{farina2006}, yielding a vanishing force for very massive quanta. These considerations point to the profound analogies between quantum and classical Casimir effects.

\section{Casimir Effects in NanoScience}

\subsection{Casimir Torques}\label{sec:torques}
\begin{figure*}[!h]
\centering
  \includegraphics[width=2\columnwidth]{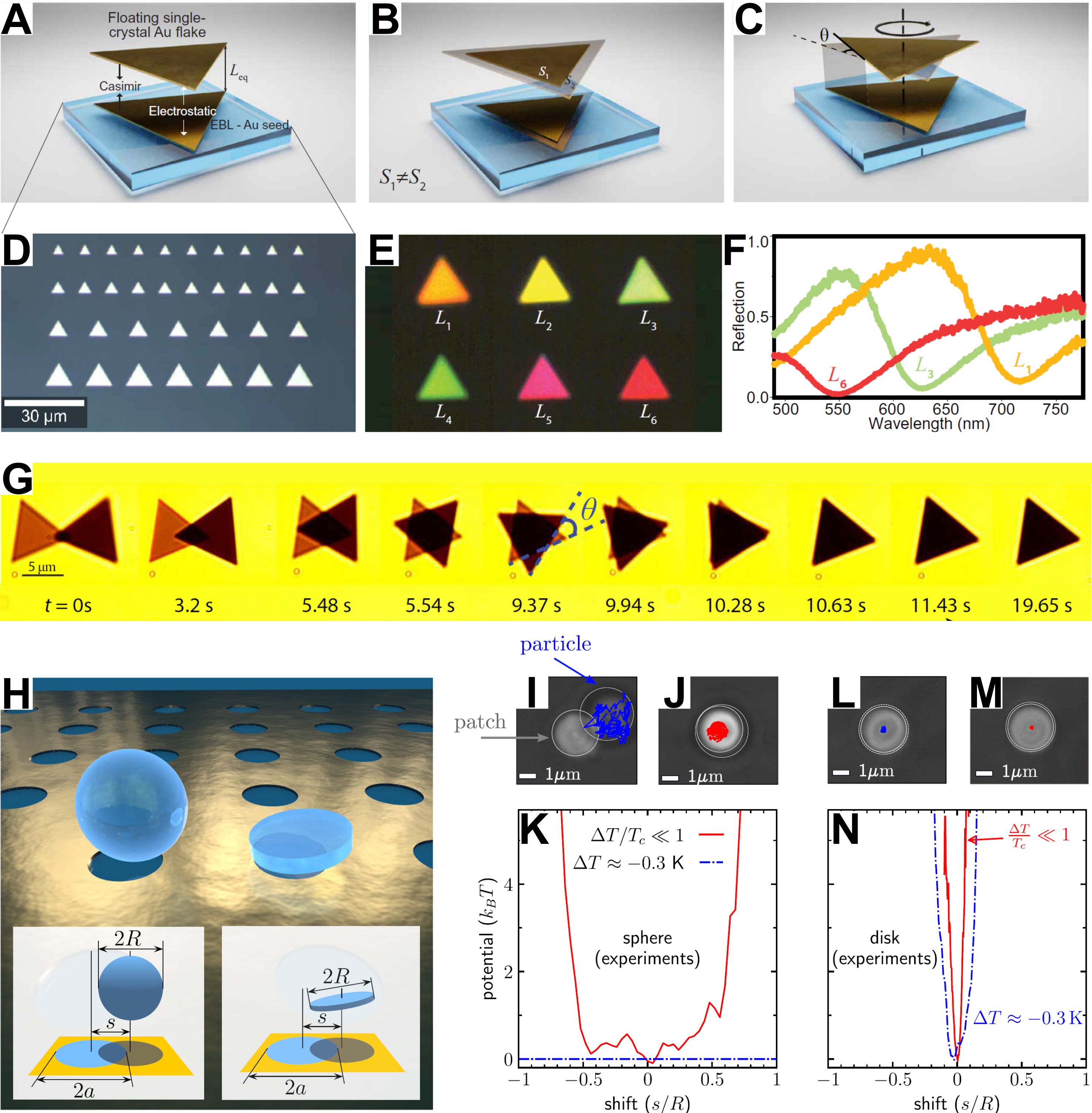}
  \caption{\textbf{Experimental systems showcasing quantum and critical Casimir torques.} 
  \textbf{(A)} Schematic illustration of the quantum Casimir assembly between a lithographically-patterned and a colloidal gold triangle. 
  \textbf{(B)} The system minimizes energy by maximizing the overlap area through rotation (quantum Casimir self-alignment).
  \textbf{(C)} During assembly, the rotation angle ($\theta$) varies. Perfect alignment corresponds to $\theta = 0$. 
  \textbf{(D)} Optical bright-field image of a glass substrate with lithographically-patterned gold equilateral triangles with $4, 5, 7, 10\,\mu m$ edge lengths.
  \textbf{(E)} True-color reflection images of quantum Casimir-aligned microcavities recorded at distinct equilibrium cavity length values ($L_i$, $i = 1-6$). 
  \textbf{(F)} Reflection spectra of aligned gold triangles, displaying their behavior as Fabry-Pérot cavities ($L_1 = 199\,nm >L_3 = 161\,nm >L_6 = 123\,nm$). 
  \textbf{(G)} Selected micrographs of the quantum Casimir-alignment process between a colloidal and a patterned gold flakes.
  \textbf{(H)} Schematic of a hydrophilic colloidal sphere (left) and disk (right) hovering above a round, hydrophilic silica patch surrounded by a hydrophobic gold layer. 
  \textbf{(I)} A colloidal sphere diffuses freely above the silica patch (blue trajectory) when the temperature of the system remains $0.3 K$ below the critical temperature, \textbf{(J)} but becomes confined at the critical temperature (red trajectory) by the critical Casimir effect.  
  \textbf{(K)} Experimentally measured potentials relative to the confinement of a colloidal sphere $0.3 K$ below the critical temperature ($\Delta T \approx-0.3\,K$) and at the critical temperature ($\Delta T/T_c\ll1$)).
  \textbf{(L)} A colloidal disk experiences confinement already $0.3 K$ below the critical temperature (blue trajectories, $\Delta T \approx-0.3\,K$). \textbf{(M)} The degree of confinement increases further at the critical temperature (red trajectories, $\Delta T/T_c\ll1$). \textbf{(N)} These observations are confirmed by experimentally measured  potentials. 
  Panels A–G are reproduced with permission from ref. \cite{Casimir-torque-Triangles}. Copyright 2024 The American Association for the Advancement of Science. Panels H–N are reproduced with permission from ref. \cite{Casimir-torque-disks}. Copyright 2024 Springer Nature Publishing Group.
  }
  \label{fgr:torques}
\end{figure*}

Measurements of the quantum Casimir force have been restricted to controlled conditions, such as dry or high-vacuum environments. The addition of a dielectric medium, such as water, allows the introduction of the Casimir interaction to the colloidal domain. Within this domain, colloidal (nano)particles of different compositions, sizes, and shapes, can serve as a fruitful testing ground for theoretical estimates for the strength and length scale of the Casimir potential, thereby enabling direct comparisons with other interactions, such as the critical Casimir force.\cite{Casimir-Quantum-Plus-Critical} 

%FIG 1 A-G
The introduction of anisotropic particles allows studying the influence of a new degree of freedom, \textit{rotation}, on the Casimir force. Under conditions of optical anisotropy, a Casimir torque develops, causing rotation of the materials under consideration to reach a position of equilibrium.\cite{Parsegian1972, Barash1978} Due to the small magnitude of the effect (torque per unit area: $\sim 10^{-9}Nm/m^2$), Casimir torques were measured only recently between a solid birefringent crystal and a liquid crystal by Somers \textit{et al.}\cite{Somers2018} In a more recent contribution, Küçüköz \textit{et al.} studied the interaction between micrometer-sized gold equilateral triangles.\cite{Casimir-torque-Triangles} The authors achieved this goal by first patterning $4-10\,\mu m$ equilateral gold triangles into a glass substrate by electron beam lithography, then studying their interaction with colloidal gold triangles in water, see Fig. \ref{fgr:torques}A and D. Theory predicts the onset of attractive quantum Casimir interactions between the gold triangles. Left unbalanced, this interaction would lead to irreversible aggregation. However, charged molecular ligands (CTAB) bound to the surfaces of the colloidal triangles contribute to a repulsive electrostatic interaction, leading to colloidally stable triangle dimers in water, Fig. \ref{fgr:torques}A. When approaching a templated triangle, a colloidal triangle tends to adopt a parallel configuration that minimizes the energy of the system by maximizing area overlap; see Fig. \ref{fgr:torques}B. As shown in Fig. \ref{fgr:torques}G, the assembly process consists of independent lateral and rotational motions of the colloidal triangle. First ($0-5.48\,s$) the colloidal triangle slides over the patterned triangle to increase overlap; then ($5.54-19.65\,s$) the colloidal triangle rotates to achieve the necessary alignment to maximize overlap. At equilibrium, the colloidal triangle minimizes rotational misalignment ($\theta=0$) while maintaining a stable separation distance $L_{eq}$, see Fig. \ref{fgr:torques}C. The authors show direct control over the value of $L_{eq}$ by varying the Debye screening length of the solution. This leads to significant changes in the color of the light reflected by the triangle dimer; see Fig. \ref{fgr:torques}E. These changes can be understood by analyzing the reflected light spectra shown in Fig. \ref{fgr:torques}F. As the triangles get closer ($L_1 = 199\,nm >L_3 = 161\,nm >L_6 = 123\,nm$), the minimum in light transmission shifts to shorter wavelengths, consistent with the behavior of each triangle dimer as a Fabry-Pérot cavity.

%FIG 1 H-N
Although the critical Casimir effect is conceptually distinct from the quantum Casimir effect, theory predicts a similar development of a torque between anisotropic particles, such as cylinders.\cite{PhysRevE.88.012137, labbe2014alignment} Recently, Wang et al. have studied the presence of critical Casimir torques in experiments.\cite{Casimir-torque-disks} The authors deposited a gold film on a glass substrate, patterning the film with round holes to achieve openings $1-2.8\,\mu m$ in size, see Fig. \ref{fgr:torques}H. They studied the critical Casimir forces between hydrophilic silica colloidal spheres ($3\,\mu m$ in diameter) or disks ($2.4\,\mu m$ in height, $0.4\,\mu m$ in height) and silica patches in a surrounding medium consisting of a water–lutidine critical mixture with a critical temperature of $T_c \approx 310\,K$. Since the critical Casimir effect ranges from attractive to repulsive depending on the wetting properties of interacting bodies, the authors controlled this parameter by making the silica patches hydrophilic by applying an oxygen plasma treatment and the gold substrate hydrophobic by functionalization with hydrophobic thiols. When the temperature of the system remains more than $0.3\,K$ away from the critical temperature, the spheres diffuse freely above the gold substrate, as shown by a typical trajectory in Fig. \ref{fgr:torques}I. When increasing the temperature to the critical value of the mixture, the sphere becomes confined to the hydrophilic silica patch, see Fig. \ref{fgr:torques}J. This is expected, as the magnitude of the critical Casimir force increases sharply close to the critical temperature. The authors extract the experimental effective potential by using the particle trajectories to evaluate the probability $P_{exp}(s)$ of a particle displacement from the center of the pattern ($s=0$) and using $U_{exp}(s)= -k_BT\ln{[P_{exp}(s)]}$. While no confining potential is observed far from the critical temperature, the sphere becomes strongly confined at the center of the silica patch at the critical temperature, with a magnitude of the confining potential exceeding $k_BT$ when the sphere strays away from the center of the patch for more than half of its radius ($s/R=0.5$), see Fig. \ref{fgr:torques}K. When using colloidal disks, the authors observed confined behavior already $0.5\,K$ away from the critical temperature. Indeed, the trajectories of the disks show confinement at the silica patch both $0.5\,K$ below the critical temperature and at the critical temperature, Fig. \ref{fgr:torques}L and M respectively. When extracting the experimental effective potential, the authors observe a sharper potential compared to the case of the spheres, suggesting a significantly reduced range of lateral movement, Fig. \ref{fgr:torques}N. The authors attribute the differences between spheres and disks to the difference in overlap area with the silica patches. Below the critical temperature, the solvent density fluctuations responsible for the critical Casimir force limit the length scale of the interaction; for this reason, only the fraction of the surface of the sphere that is closest to patch is sensitive to the critical Casimir attraction/repulsion to the hydrophilic/hydrophobic portions of the substrate, while the entire bottom surface of the disk has the same sensitivity.

%%%%%%%%%%%%%%% CASIMIR RESONATORS #1 %%%%%%%%%%%%%%%

\subsection{Solid-State-based Casimir assembly and resulting applications}\label{sec:SSCas}
\begin{figure*}[!h]
\centering
  \includegraphics[width=2\columnwidth]{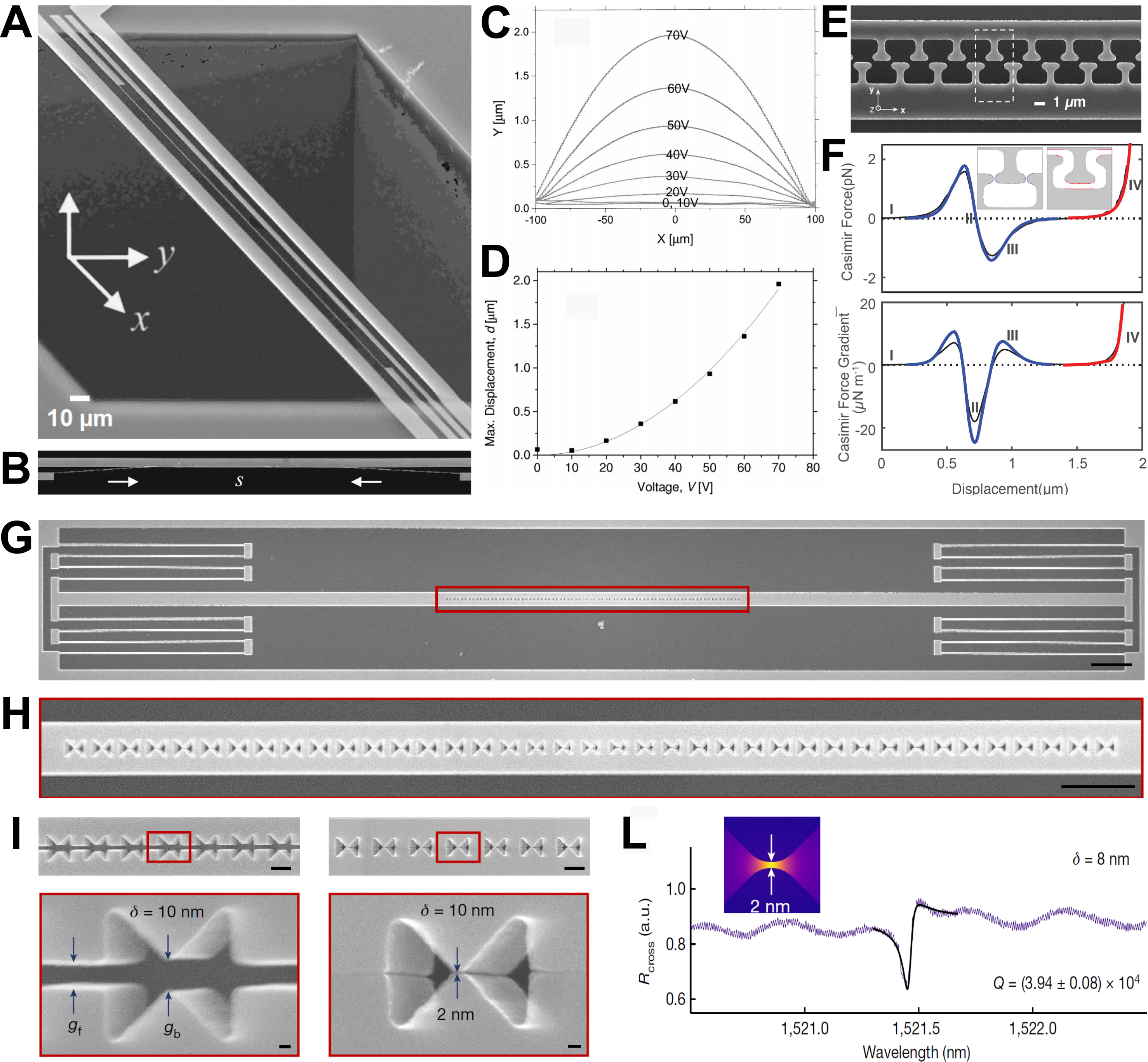}
  \caption{\textbf{Solid-state optical resonators based on the quantum Casimir effect. }
  \textbf{(A)} A scanning electron micrograph of a $200\,\mu m\times 0.25\mu m$ gold beam suspended between two electrodes.
  \textbf{(B)} Quantum Casimir forces induce surface adhesion between beam and electrode. 
  \textbf{(C)} Measured deflection of the beam due to application of electrostatic force.
  \textbf{(D)} Displacement of the center of the beam as a function of applied voltage.
  \textbf{(E)} Top-view scanning electron micrograph of two arrays of T-shaped protrusions at rest (zero displacement). The dashed rectangle illustrates one unit cell. Scale bar illustrates $1\,\mu m$. 
  \textbf{(F)} Calculated Casimir force and force gradient per unit cell along the y axis as a function of displacement. The red/blue lines represent the Casimir force calculated using the Derjaguin approximation for regions outlined in red/blue in the right/left inset.\cite{Derjaguin} The black line represents boundary-element method calculations.\cite{BEM}.
  (G) Scanning electron micrograph of the full device, including the beams (center) and spring suspensions (sides). Scale bar indicates $2\,\mu m$.
  (H) Scanning electron micrograph highlighting the bowtie-shaped cavities. Scale bar indicates $1\,\mu m$.
  (I) (Left) Scanning electron micrograph of the central part of the cavity before device collapse. Initial gap between the beams: $g_f = 50\,nm$. Initial gap between the bowties: $g_b = 52\,nm$. (Right) Scanning electron micrograph of the central part of a nanobeam cavity after device collapse, with the approximately $2\,nm$ gap indicated in the zoom-in.
  (L) Representative cross-polarized scattering spectrum (purple line) of a cavity. The fitted lineshape over the fitting range is included (black line). Inset: Absolute value of the electric field of the nanobeam cavity with a linear colour map.
  Panels A–D are reproduced with permission from ref. \cite{Casimir-gold-beam}. Copyright 2001 American Physical Society. Panels E–F are reproduced with permission from ref. \cite{Casimir-tunable-metamaterial}. Copyright 2017 Springer Nature Publishing Group. Panels G–J are reproduced with permission from ref. \cite{Casimir-photonic-cavities}. Copyright 2023 Springer Nature Publishing Group.
  }
  \label{fgr:resonator-1}
\end{figure*}

%FIG 2 A-D
Achieving direct control over Casimir forces - whether quantum or critical - provides unique pathways to the fabrication and actuation of opto-electro-mechanical devices operating on the micro- and nanoscale.\cite{Chan2001} Theses devices typically consist of resonators that function as sensors or actuators, with a physical response mechanism that can be reversible or irreversible. The main practical limitation of this emerging technology is \textit{stiction}: the adhesion or collapse of moving parts due to irreversible attraction induced by quantum Casimir or capillary forces. Estimating the magnitude of these effects is not straightforward. Buks \& Roukes estimated the adhesion energy of gold on gold by microfabricating the gold nanobeam shown in Fig. \ref{fgr:resonator-1}A.\cite{Casimir-gold-beam} The authors apply an electric field to the beam, studying the onset of nanobeam adhesion as shown in Fig. \ref{fgr:resonator-1}B. As the magnitude of the electric field increases, the nanobeam profile along the x-y plane begins to change, bending by up to $2\,\mu m$ at $70\,V$, see Fig. \ref{fgr:resonator-1}C-D. The authors study the mechanics of the bending beam, extracting an adhesion energy of $\gamma = 0.066\,J/m^2$. Importantly, this value is six times lower than the expected value for Casimir-induced adhesion between atomically flat surfaces, $\gamma_{Casimir} = 0.4\,J/m^2$. The discrepancy between adhesion values measured in experiments and expected from theory is likely a consequence of a hard reality-check from surface science: the presence of organic residues from microfabrication and/or non-atomically flat surfaces would decrease the estimated values for quantum Casimir adhesion significantly.      

%Fig 2 E-F
Despite the difficulty in measuring the magnitude of Casimir interactions, their nature depends on physical parameters of the system, such as the chemical composition and relative affinity of the interacting surfaces and the medium confined between them. For these reasons, the magnitude and particularly the sign of Casimir interactions are not easily changed on the fly. Recently, Tang \textit{et al.} microfabricated a clever design allowing to change the sign of the quantum Casimir interaction at will.\cite{Casimir-tunable-metamaterial} The device shown in Fig. \ref{fgr:resonator-1}E consists of two flat silicon surfaces endowed with $\sim 2\,\mu m$-sized intercalated \textit{T}-shaped protrusions. Effectively, this system behaves as a spring based on the Casimir force. As the distance between the two silicon surfaces decreases along the y-direction, the y-component of the Casimir interaction between the lateral surfaces of the protrusions increases, leading to a net attractive interaction as shown in Fig. \ref{fgr:resonator-1}F, I. Once the top of the \textit{T}-shaped protrusions align, the y-component of the Casimir force along the axes goes to zero, turning off the Casimir interaction as shown in Fig. \ref{fgr:resonator-1}F, II. When increasing the displacement further, the Casimir force along the y-direction acts as a restoring force, effectively changing sign and leading to a net repulsive interaction, Fig. \ref{fgr:resonator-1}F, III. For the largest displacements, the Casimir interaction between the top of the protrusions and the silicon surface dominates, leading once again to a net attraction, Fig. \ref{fgr:resonator-1}F, IV. When computing the force gradient for the different configurations, corresponding to the spring constant of the system, this design shows a vanishing stiffness at zero displacement (Fig. \ref{fgr:resonator-1}F, I), followed by an increase to $\sim10\,\mu N/m$, and notably a sign change to reach $\sim-20\,\mu N/m$ (Fig. \ref{fgr:resonator-1}F, II), elegantly displaying the transition between attractive and repulsive regimes.  

Similar structural instabilities were recently used by Babar \textit{et al.} to fabricate functional devices, such as photonic cavities.\cite{Casimir-photonic-cavities} The authors built a silicon-on-insulator device based on two suspended beams connected to mechanical springs, as shown in Fig. \ref{fgr:resonator-1}G-H. When the gap between the beams is sufficiently wide, the restoring force of the springs is able to maintain the beams separated; however, when the gap becomes too narrow, the restoring force cannot keep the beams apart, resulting in their mechanical collapse and adhesion by the Casimir force. During device fabrication, the authors etched the beams with triangular features, such that the structural collapse would result in the bowtie-shaped cavities shown in Fig. \ref{fgr:resonator-1}I, leaving a $2\,nm$ air gap at the center. Electromagnetic field simulations predict that this air gap should behave as a photonic resonator, efficiently confining light in a spectral window ideal for telecommunications, see the inset in Fig. \ref{fgr:resonator-1}L. The authors prove this claim by characterizing the nanocavities using far-field resonant scattering. The position ($\lambda = 1521.5\,nm$) and width ($\Delta\lambda$) of the resonance shown in Fig. \ref{fgr:resonator-1}L imply a  quality factor of $Q = \lambda/\Delta\lambda = 3.9\times10^4$, a competitive figure of merit in such a valuable spectral range.

%%%%%%%%%%%%%%% CASIMIR RESONATORS #2 %%%%%%%%%%%%%%%
\subsection{Colloidal-State-based Casimir assembly and resulting applications}\label{sec:CollCas}
\begin{figure*}[!h]
\centering
  \includegraphics[width=2\columnwidth]{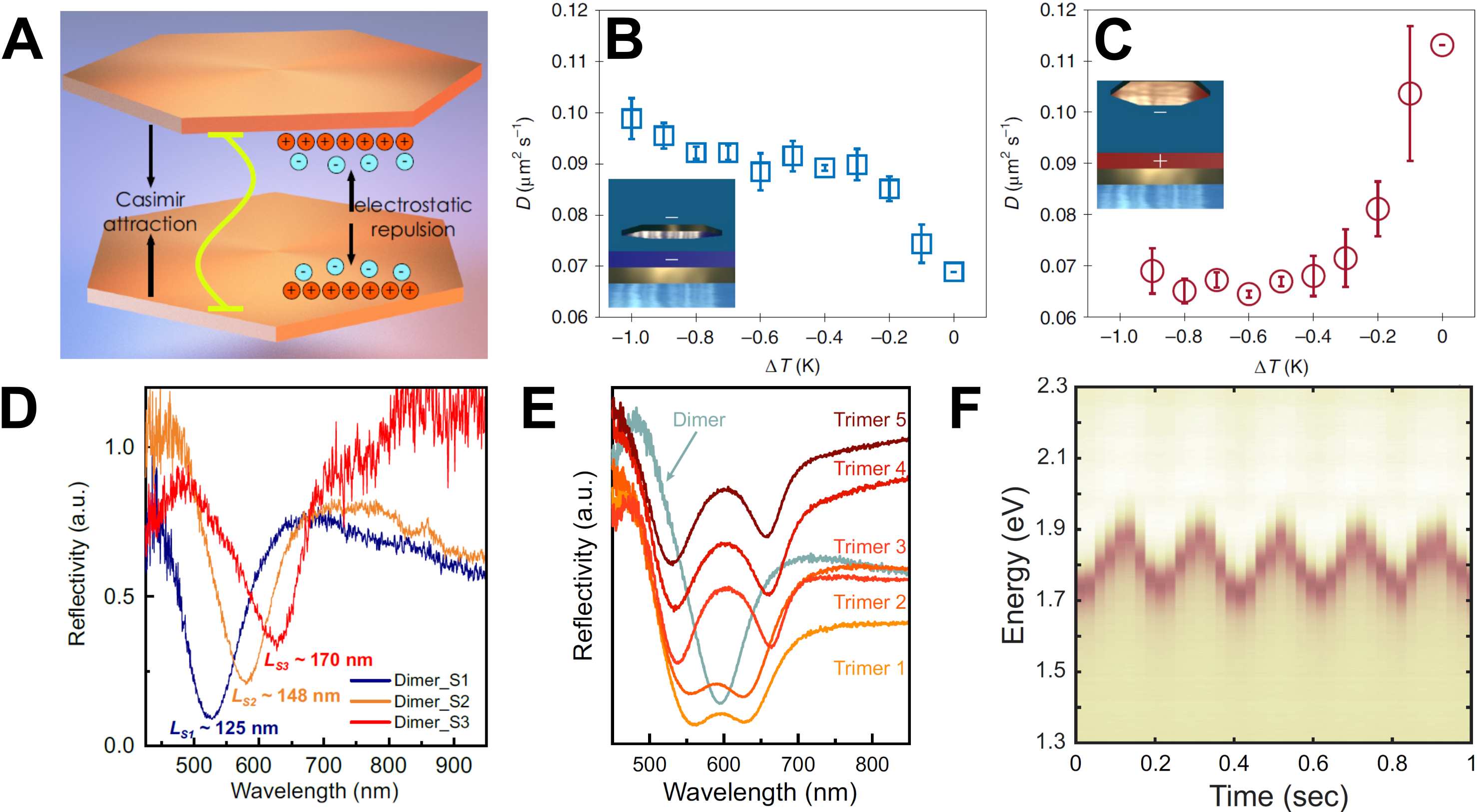}
  \caption{\textbf{Colloidal-state based optical resonators based on the quantum and critical Casimir effects.}
  \textbf{(A)} Schematic of hydrophilic gold flakes subjected to quantum and critical Casimir interactions.
  \textbf{(B)} The measured diffusion coefficient of a hydrophilic hexagonal gold flake hovering above a hydrophilic surface (symmetric boundary conditions) decreases as the
temperature approaches the critical temperature. 
  \textbf{(C)} The measured diffusion coefficient of a hydrophilic hexagonal flake hovering above a hydrophobic surface (antisymmetric boundary conditions) increases upon approaching the critical temperature.
  \textbf{(D)} Quasi-normal incidence reflectivity spectra of dimer cavities assembled with variable cavity lengths ($L_{S}$) by varying the free ligand concentration.
  \textbf{(E)} Quasi-normal incidence reflectivity spectra of self-assembled trimer cavities. The grey-blue curve shows the reflectivity spectrum of a dimer.
\textbf{(F)} Time-resolved reflectivity spectra for a Fabry-Pérot cavity under optical modulation by $5\,Hz$ pulsed laser irradiation.
Panels A, D–F are reproduced with permission from ref. \cite{Casimir-tunable-resonators}. Copyright 2021 Springer Nature Publishing Group. Panels B and C are reproduced with permission from ref. \cite{Casimir-Quantum-Plus-Critical}. Copyright 2022 Springer Nature Publishing Group.
  }
  \label{fgr:resonator-2}
\end{figure*}

The transition from the solid- to the colloidal-state offers both advantages and disadvantages. While microfabrication techniques offer a large degree of control over the horizontal plane, colloidal assembly has the potential to extend device fabrication to three dimensions. Recently, Schmidt \textit{et al.} paved the way in this fascinating direction by exploiting both quantum \textit{and} critical Casimir forces.\cite{Casimir-Quantum-Plus-Critical} In their work, the authors used hexagonal gold flakes stabilized in a near-critical mixture of water and 2,6-dimethylpyridine by hydrophilic CTAB ligands, as shown in Fig. \ref{fgr:resonator-2}A. When using a hydrophilic silica substrate, the diffusion coefficient of the flakes measured in proximity of the substrate decreases as a function of temperature as shown in Fig. \ref{fgr:resonator-2}B. As the temperature of the mixture approaches the critical value, solvent density fluctuations arise, causing the onset of critical Casimir interactions. The symmetric (hydrophilic flakes - hydrophilic substrate) boundary conditions favor attractive interactions, leading to slower diffusion as the flake is pulled closer to the substrate from $100\,nm$ away from criticality to $~70\,nm$ near-criticality. Using a hydrophobic gold substrate results in the opposite effect, as shown in Fig. \ref{fgr:resonator-2}C. As the temperature of the mixture increases, the authors observed an increase in the diffusion coefficient of the flakes close to the substrate, indicative of the onset of a repulsive force. This is consistent with theoretical expectations, as asymmetric boundary conditions favor repulsive critical Casimir interactions, leading to faster diffusion as the flake is pushed to an equilibrium distance of $200\,nm$ away from the substrate at near-criticality.  

The interplay between electrostatic and Casimir forces can be exploited to build optical resonators based on colloidal particles. When relying on critical Casimir forces, the structural stability and optical properties of the resonators are sensitive to the temperature of the solvent mixture, allowing for directed, tunable, and reversible assembly.\cite{Casimir-Quantum-Plus-Critical} While quantum Casimir forces provide a less easily tunable interaction, they favor the structural persistence of assemblies, a desirable property for functional devices. Nevertheless, structural tunability can still be achieved by modulating the repulsive electrostatic interaction, as recently shown by Munkhbat \textit{et al.}\cite{Casimir-tunable-resonators} The authors observed that gold nanoflakes dispersed in an aqueous solution of CTAB ($2\,mM$) tend to form stable dimers. These dimers effectively behave as Fabry-Pérot cavities, confining light of a single wavelength. In experiment, this behavior results in a pronounced dip in optical reflectivity data, as shown in Fig. \ref{fgr:resonator-2}D. Decreasing the concentration of free CTAB ligands in solution increases the Debye screening length, increasing both the range and the magnitude of the electrostatic repulsion between gold flakes. Indeed, decreasing the CTAB concentration 2- and 3-fold results in an increase in the surface-to-surface distance between gold flakes in a dimer from $125\,nm$ to $148\,nm$ and $170\,nm$. As a consequence, the spectral position of the reflectivity minimum redshifts accordingly. Furthermore, increasing the complexity of the assembled resonators brings interesting developments. Assembling trimers results in an optical resonator characterized by two minima in reflectivity, as shown in Fig. \ref{fgr:resonator-2}E. The authors show that this emergent optical feature is the result of the coupling between the optical modes confined between the two pairs of dimers, resulting in symmetric and anti-symmetric collective modes of the trimer. While controlling the optical properties of optical resonators by modulating their structure is certainly valuable, active optical tuning would unlock a different set of applications towards sensing and communications. The authors demonstrate potential in this direction by showing that laser excitation induces sufficient light pressure to increase the cavity length. Once the excitation is turned off, the equilibrium value of the cavity length is quickly restored by the quantum Casimir attraction. Repeating the experiment using a pulsed excitation source results in a reversible shift of $0.2\,eV$, corresponding to $20\,nm$.

\subsection{Casimir Forces Tunable by Dielectric Control}\label{sec:TunCasDie}

Recently, researchers have begun exploring the use of external triggers to modify interactions. For instance, scientists have started exploring the possibility of using an external quantum or classical fields to control the quantum Casimir interaction, enhancing or suppressing its magnitude, as well as the possibility of switching from attractive to repulsive behavior.\cite{Thirunamachandran80,MilonniSmith96,Sherkunov09,FiscelliRizzuto20,JakubecSolano24} As we have mentioned, both the magnitude and the sign of the Casimir interaction depend on various characteristics of the system, and as such the magnitude of the effect cannot be easily modified. There is, however, a strong motivation to develop new ways to control this interaction at will. Apart from those ideas related to the fabrication of functional devices, such as micro- and nano- electromechanical devices, other opportunities have emerged through \textit{superlubricity}.\cite{iannuzzi2007ultra, Feiler2008} Indeed, changing the sign of the quantum Casimir force may lead to a significant decrease in friction between any two bodies, with the potential to increase the energetic efficiency of most processes that shape our world.\cite{Hod2018} In the extreme case of zero friction, we may be able to obtain quantum levitation,\cite{PhysRevB.89.201407, Spreng:24} which may represent a new power source for propulsion in space. 

While we are still far from achieving these goals, the last two decades have seen significant advances. First, Munday \textit{et al.} have demonstrated experimentally the feasibility of achieving repulsive quantum Casimir forces between two materials (1 and 2) separated by a medium (3).\cite{Munday2009} The authors recognized that for a repulsive Casimir force to arise, the inequality 

\begin{eqnarray}
\label{eq:Cas-rep}
-(\epsilon_1-\epsilon_3)(\epsilon_2-\epsilon_3)>0
\end{eqnarray}

must be satisfied, where $\epsilon_i$ represents the dielectric response function of the $i$-th component. The presence of symmetric boundary conditions ($\epsilon_1=\epsilon_2$) always leads to an attractive quantum Casimir interaction since eq. \ref{eq:Cas-rep} yields a negative result, $-(\epsilon_1-\epsilon_3)(\epsilon_2-\epsilon_3) = -(\epsilon_1-\epsilon_3)^2<0$. Therefore, the adoption of asymmetric boundary conditions ($\epsilon_1\neq\epsilon_2$) is imperative for the development of repulsive quantum Casimir interactions. In particular, the choice of components must satisfy the inequality 

\begin{eqnarray}
\label{eq:Cas-rep-2}
\epsilon_1>\epsilon_3>\epsilon_2
\end{eqnarray}

The authors tested this prediction using gold and silica as materials, and bromobenzene as medium, one of the rare combinations that satisfies eq. \ref{eq:Cas-rep-2} since $\epsilon_{\mbox{\small{Gold}}}>\epsilon_{\mbox{\small{Bromobenzene}}}>\epsilon_{\mbox{\small{Silica}}}$ over a large range of imaginary frequencies ($<10^{16} \,rad/s$). To prove their hypothesis, the authors used atomic force microscopy to measure repulsive interactions between a gold sphere and a silica plate at distances larger than $20\,nm$, corresponding to the low-frequency domain where eq. \ref{eq:Cas-rep-2} holds. Since Casimir forces at large separations mainly result from low frequencies, and those at small separations from high frequencies, a stable Casimir equilibrium may be realized if the dielectric response of a physical system only contributed attractive forces at small frequencies and repulsive forces at large frequencies. This possibility had been theoretically predicted\cite{PhysRevLett.104.160402} but never realized experimentally until the work of Zhao \textit{et al.} in developing stable Casimir equilibria in a teflon-ethanol-gold system.\cite{Zhao2019} 

%%%%%%%%%%%%%%% CASIMIR MAGNETICS %%%%%%%%%%%%%%%
\subsection{Casimir Forces Tunable by External Magnetic Fields}\label{sec:TunCasMag}

\begin{figure*}[!h]
\centering
  \includegraphics[width=2\columnwidth]{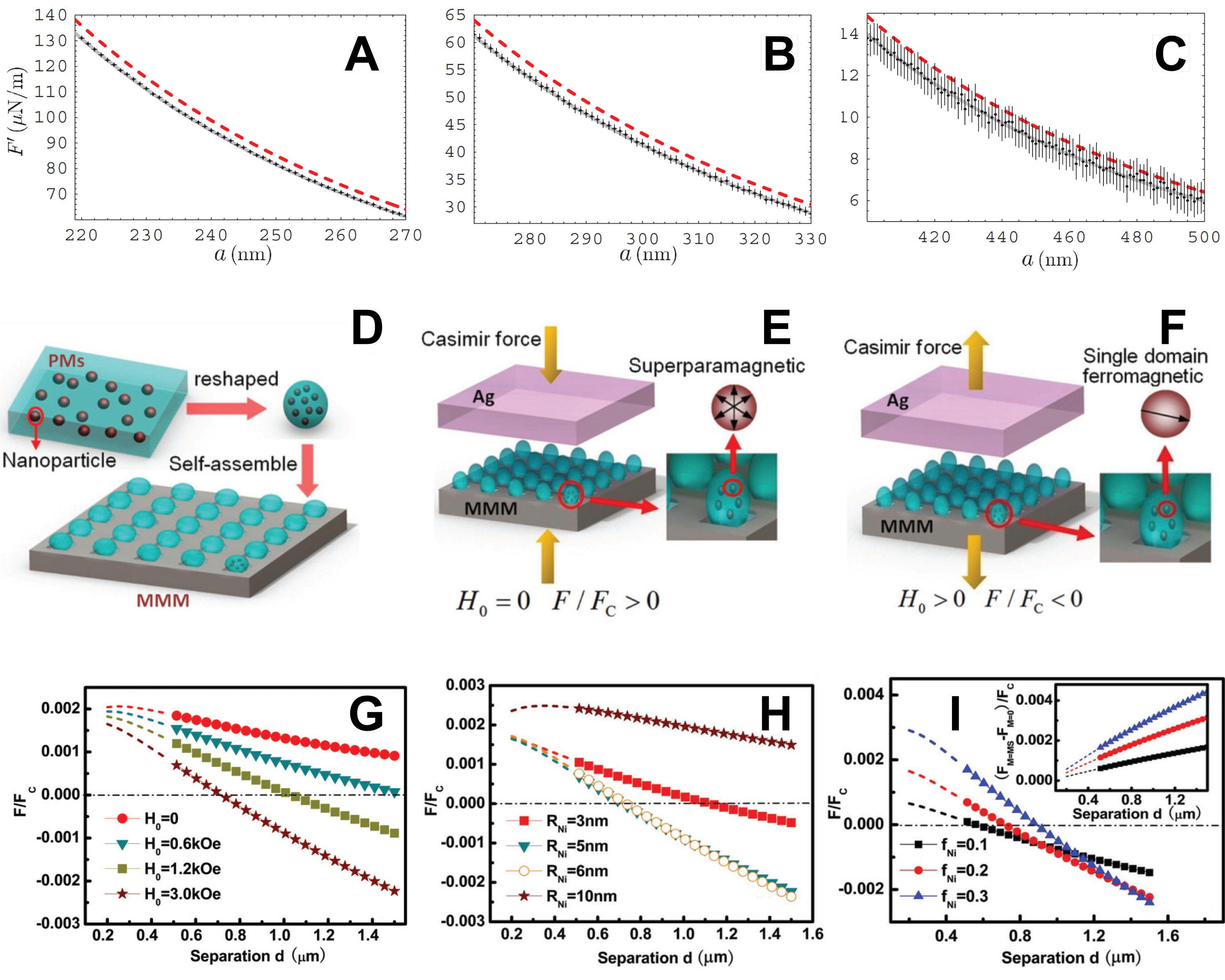}
  \caption{\textbf{Tuning quantum Casimir forces by magnetic fields.}
  \textbf{(A-C)} Comparison for the gradient of the Casimir force between experiment (crosses) and theory (solid and dashed bands computed including or excluding magnetic properties, respectively).
  \textbf{(D)} Schematic of the construction of a magnetic metamaterial plate.
  \textbf{(E)} Schematic depicting the interaction between a silver plate and the magnetic metamaterial plate in the absence of an external magnetic field ($H_0 = 0$), resulting in attractive Casimir forces ($F/F_C>0$).
  \textbf{(F)} Schematic depicting the interaction between a silver plate and the magnetic metamaterial plate in the presence of an external magnetic field ($H_0 > 0$), resulting in repulsive Casimir forces ($F/F_C<0$).
  \textbf{(G)} Computed normalized Casimir force between $R_{Ni} = 5\,nm$ Nickel nanoparticle-based magnetic metamaterial plate and a silver plate under different magnetic field intensities at a Nickel volume fraction of $f_{Ni}=10\%$.
  \textbf{(H)} Computed normalized Casimir force between Nickel nanoparticle-based magnetic metamaterial plate and a silver plate under $H_0 = 3.0\,kOe$ magnetic field for different radii of Nickel nanoparticles, $R_{Ni}$.
  \textbf{(I)} Computed normalized Casimir force between $R_{Ni} = 5\,nm$ Nickel nanoparticle-based magnetic metamaterial plate and a silver plate under $H_0 = 3.0\,kOe$ magnetic field for different Nickel volume fractions $f_{Ni}=10\%-30\%$. Inset shows the adjusting ranges with different $f_{Ni}$.
  Panels A–C are reproduced with permission from ref. Casimir-magnetic-tuning. Copyright 2012 American Physical Society. Panels D–I are reproduced with permission from ref. \cite{Casimir-magnetic-metamaterial}. Copyright 2014 American Physical Society.
  }
  \label{fgr:magnetics}
\end{figure*}

Apart from dielectric control, using an external magnetic field to drive a change in the magnitude of the quantum Casimir interaction has provided a new direction. In their contribution, Banishev \textit{et al.} experimentally studied the effect of a magnetic field on the magnitude of the quantum Casimir force between gold-nichel,\cite{Casimir-magnetic-tuning} nichel-nichel,\cite{Casimir-magnetic-tuning-Ni}, and gold-gold coated surfaces.\cite{Casimir-magnetic-tuning-Au} Nichel, a ferromagnetic metal, provides the magnetic field. The magnitude of the quantum Casimir force is measured by frequency modulation atomic-force microscopy, where the tip is forced to oscillate near its resonant frequency such that any changes in this frequency are proportional to the gradient of the force. The results in Fig. \ref{fgr:magnetics}A-C show the gradient of the quantum Casimir force ($F^\prime = dF/da$) between gold-coated sphere and nichel-coated plate as a function of the separation distance ($a = 220-500\,nm$). As expected, when the separation distance decreases, the experimentally-measured gradient of the quantum Casimir force increases (crosses). The authors offer a comparison with theory including (full lines) and omitting (dashed lines) the presence of a magnetic field. The small relative experimental error confirm the excellent agreement with theory when including the effect of a magnetic field, and reject the same theoretical approach when excluding the effect of a magnetic field. While the care devoted by the authors in performing such challenging measurements deserve praise and recognition, the magnetic-field induced changes to the gradient of the quantum Casimir force are admittedly still small. For instance, at a separation distance of $a=250\,nm$, the magnetic field induces a decrease in the gradient of the quantum Casimir force of only $\Delta F^\prime/F^\prime\approx4\%$, still far from achieving the reversal of the force from attractive to repulsive in this experimental configuration.

Theoretical predictions show that the choice of a different experimental setup, such as a magneto-dielectric material consisting of a ferromagnetic guest embedded within a dielectric host, may lead to more encouraging results.\cite{Casimir-magnetic-tuning-EMM-2, Casimir-magnetic-tuning-EMM} In their contribution, Ma \textit{et al.} come closer to achieving force reversal between plates by using a superparamagnetic metamaterial.\cite{Casimir-magnetic-metamaterial} Such metamaterial is constructed as shown in Fig. \ref{fgr:magnetics}D. First, superparamagnetic Nichel nanoparticles are dispersed in a polystyrene matrix. This matrix is then reshaped into spheres and placed on a substrate to form the final metamaterial. This metamaterial is then allowed to interact with a silver plate in the absence or in the presence of an external magnetic field, Fig. \ref{fgr:magnetics}E and F respectively. 

In the absence of an external magnetic field, the Casimir force between the metamaterial and silver plates is attractive, as shown in Fig. \ref{fgr:magnetics}G. The application of an increasingly strong magnetic field causes the Casimir force at $d>0.7\,nm$ separations to turn from attractive to repulsive, as can be seen in the change of sign of the normalized Casimir force $F/F_C$, where $F_C = -hc\pi^2/240d^4$ is the Casimir force between two conductive plates. The application of a magnetic field causes the magnetic moments of the nanoparticles to align, increasing the value of the magnetic permeability above that of the permittivity in the low imaginary frequency domain ($\xi < 10^{14}\,rad/s$), resulting at sufficiently large separations in a repulsive Casimir forces. The main limitation of this approach lies with the choice of nanoparticle size and polydispersity,\cite{Casimir-magnetic-metamaterial-PD} as shown in Fig. \ref{fgr:magnetics}H. Decreasing the radius of the nanoparticles below $5\,nm$ decreases the range over which the interactions are repulsive, as the magnetization of each particle decreases. Increasing the radius of the nanoparticles to $10\,nm$ results in fully attractive interactions, as the magnetic response shifts to lower frequencies and hence larger separations. Increasing the Nichel volume fraction $f_{Ni}$ allows direct tuning of the transition point from attraction to repulsion from $d\approx0.6\,\mu m$ ($f_{Ni} = 10\%$) to $d\approx0.9\,\mu m$ ($f_{Ni} = 30\%$). 

Recently, Zhang \textit{et al.} took advantage of these theoretical considerations to achieve magnetic-field tuning of the Casimir force in experiments.\cite{Casimir-magnetic-experiment} The authors used an aqueous dispersion of superparamagnetic iron oxide nanoparticles as medium to mediate the Casimir interaction between gold and silica surfaces. The magnetization of the ferrofluid induces the onset of repulsive Casimir interactions for $d>18\,nm$, where the transition distance is tunable by the magnitude of the applied field and nanoparticle concentration. 

\section{Conclusions and Outlook}
We have highlighted the significance of fluctuation-induced forces by exploring the fundamental principles of quantum and critical Casimir effects and their role in modern technology. Quantum and critical Casimir effects share a common mathematical background, resulting in several similarities in their dependence on geometric confinement, boundary conditions, and material properties. However, the different physical origins of the quantum (quantum electrodynamics) and critical (critical phenomena) Casimir effects result in distinctive and orthogonal characteristics. We have shed light on these characteristics, highlighting the roles of Casimir \textit{torques} in achieving self-alignment and Casimir \textit{forces} in achieving self-assembly of functional materials. The interplay between quantum fluctuations of the electromagnetic field and classical fluctuations of a near-critical solvent offers a rich playground for emergent interactions at the multi-dimensional crossroads between experiment and theory, soft and hard condensed matter, physics and chemistry. 

Beyond fundamental science, the quantum and critical Casimir effects promise technological advances in photonics and opto-mechanics towards the development of micro-electro-mechanical systems. So far, Casimir forces have enabled the fabrication of switchable physical structures able to trap light on length scales smaller than the wavelength. However, recent advances in superlubricity and dynamic force tuning suggest a path toward the assembly of more complex devices capable of converting electromagnetic and thermal energy in mechanical energy. The intrinsic character of the Casimir effect, and the possibility of tuning the magnitude, length scale, and sign of the force by changing the dielectric or thermal properties of materials may enable valuable applications. For instance, because of the omnipresent character of the interaction, space travel may benefit from the development of a propulsion engine based on the Casimir effect. The high stakes at play have motivated significant research efforts towards the use of other orthogonal means to control the Casimir interactions, such as through magnetic fields.

Looking forward, we believe that the most significant challenge ahead consists in achieving real-time, full tunability of Casimir interactions by the application of an external trigger. Furthermore, understanding the non-additive interactions resulting from large-scale fluctuations may allow the integration of Casimir forces in modern nanofabrication techniques. Overall, the convergence of theoretical insights and experimental progress will be key to unlocking the full potential of these fluctuation-induced forces in next-generation technologies.

\section*{Author Contributions}
The manuscript was written through the contributions of all authors. All authors have given approval to the final version of the manuscript.

\section*{Conflicts of interest}
There are no conflicts to declare.

\section*{Data availability}
No new data were generated or analysed as part of this review.

\section*{Acknowledgements}
R. P. and E. M. acknowledge funding received through “UNIPA – Misura B del Piano di azioni VQR”. E. M., R. P. and
L. R. acknowledge support from the PNRR MUR Project MINTQT, Partenariato Esteso NQSTI PE00000023, Spoke 9 –
CUP: E63C22002180006. E. M. acknowledges travel support from the National Science Foundation under the IMOD
Integrative Travel Program Award, grant no. DMR-2019444, and from the European Commission – Horizon Europe – Next
Generation Internet Enrichers programme, grant agreement no. 101070125. E. M. is grateful to the European Union –
NextGenerationEU – funding MUR D. M. 737/2021 for funding his position at Unipa. The authors acknowledge funding
received from Unipa through “Fondo Finalizzato alla Ricerca di Ateneo” (FFR) 2022–2024 (E. M.), 2023–2024 (L. R.), 2023 (R. P.).

%%%END OF MAIN TEXT%%%

% The \balance command can be used to balance the columns on the final page if desired. It should be placed anywhere within the first column of the last page.

% \balance

% If notes are included in your references you can change the title from 'References' to 'Notes and references' using the following command:
% \renewcommand\refname{Notes and references}

%%%REFERENCES%%%

\bibliography{main} 
\bibliographystyle{rsc} %the RSC's .bst file
\end{document}